\documentclass[a4paper]{elsarticle}

\usepackage{amssymb}
\usepackage{tabularx}
\usepackage{tablefootnote}
\usepackage{float}
\usepackage{subfloat}
\usepackage{listings}
\usepackage{hyperref}

\restylefloat{table}

\newcommand{\swig}{\texttt{swig}}
\newcommand{\SWIGCommon}{\texttt{SWIG\-Common.i}}
\newcommand{\SWIGOctave}{\texttt{SWIG\-Octave.i}}
\newcommand{\SWIGPython}{\texttt{SWIG\-Python.i}}
\newcommand{\generateswigiface}{\texttt{generate\-\_swig\-\_iface.py}}

\journal{SoftwareX}

\begin{document}

\begin{frontmatter}

\title{SWIGLAL: Python and Octave interfaces to the LALSuite gravitational-wave data analysis libraries}

\author[anu,aei]{Karl Wette}
\address[anu]{ARC Centre of Excellence for Gravitational Wave Discovery (OzGrav) and Centre for Gravitational Astrophysics, Australian National University, Canberra ACT 2600, Australia}
\address[aei]{Max Planck Institute for Gravitational Physics (Albert Einstein Institute), D-30167 Hannover, Germany}
\ead{karl.wette@anu.edu.au}

\begin{abstract}
  The LALSuite data analysis libraries, written in C, implement key routines critical to the successful detection of gravitational waves, such as the template waveforms describing the merger of two black holes or two neutron stars.
  SWIGLAL is a component of LALSuite which provides interfaces for Python and Octave, making LALSuite routines accessible directly from scripts written in those languages.
  It has enabled modern gravitational-wave data analysis software, used in the first detection of gravitational waves, to be written in Python, thereby benefiting from its ease of development and rich feature set, while still having access to the computational speed and scientific trustworthiness of the routines provided by LALSuite.
\end{abstract}


\end{frontmatter}

\section*{Required Metadata}\label{sec:required-metadata}

\noindent
\begin{tabularx}{\textwidth}{|l|X|X|}
  \hline
  \textbf{Nr.} & \textbf{Code metadata description}                               & \textbf{Please fill in this column}                                      \\
  \hline
  C1           & Current code version                                             & 6.21.0                                                                   \\
  \hline
  C2           & Permanent link to code/repository used for this code version     & \url{https://github.com/kwwette/swiglal}                                 \\
  \hline
  C4           & Legal Code License                                               & GPL-2.0                                                                  \\
  \hline
  C5           & Code versioning system used                                      & git                                                                      \\
  \hline
  C6           & Software code languages, tools, and services used                & C with SWIG directives, Python, Octave, LALSuite dependencies            \\
  \hline
  C7           & Compilation requirements, operating environments \& dependencies & SWIG $\ge$ 2.0.12, Python $\ge$ 2.6, NumPy $\ge$ 1.3, Octave $\ge$ 3.2.0 \\
  \hline
  C8           & If available Link to developer documentation/manual              & See \url{https://github.com/kwwette/swiglal/blob/master/README.md}       \\
  \hline
  C9           & Support email for questions                                      & See \url{https://github.com/kwwette/swiglal/blob/master/README.md}       \\
  \hline
\end{tabularx}

\section{Motivation and Significance}\label{sec:motiv-sign}

The choice of programming language is a critical decision in the design of scientific software.
Languages such as C provide a low level of abstraction between the programmer and the machine architecture, and are compiled to machine code for best performance.
The lack of abstraction, however, places a higher burden on the developer to manually handle low-level tasks, such as memory management, which detracts from the scientific problem at hand.
High-level scripting languages, of which Python~\cite{python} and Octave~\cite{octave} are two examples, provide a higher level of abstraction from the machine architecture, freeing the developer to focus on the algorithm, reducing development time, and facilitating the rapid prototyping of new ideas.
They also provide a richer set of features, either built into the language or else available through easy-to-install packages downloaded from a central repository.
They are generally not compiled to machine code, however, and therefore performance may not match that provided by low-level machine code.

Often, a new software package will want to make use of existing libraries which provide routines which are particularly efficient, well-tested and trusted by the wider scientific community, and/or difficult to re-implement.
In such cases, the developer may be constrained to use a particular language -- the same language as the existing library -- and therefore be forced to accept the costs and benefits of that particular language.
A solution to this problem is to write a software wrapper around the existing library, which then exposes its routines so that it can be used from the programming language of choice.
For example, software wrappers can enable the developer to make use of libraries written in C, while also benefiting from the ease of development and rich feature set provided by high-level languages such as Python.

The first detections of gravitational waves, from the merger of two black holes~\cite{LIGOVirg2016:ObsGrvWvBnBHMr} and from two neutron stars~\cite{LIGOVirg2017:GWObsGrvWBNtSIn}, were made possible through, amid many other advances, the careful implementation and rigorous testing of data analysis software.
LALSuite (LSC Algorithm Library Suite;~\cite{lalsuite}) is a collection of software routines for gravitational-wave data analysis, written in C, and developed since 2000.
As of version 6.67~\cite{lalsuite-6.67}, LALSuite provides, along with $\sim 230$ executables, 9 libraries which collectively export a large number of symbols, and represents a significant code base of hundreds of thousands of lines of C code (Table~\ref{tab:swiglal-exports}).
It provides atomic data types for fixed-width integer, floating-point, and complex numbers; and compound data types called ``structs'', accompanied by functions which create, destroy, and manipulate them.
(Structs in C are conceptually equivalent to classes in Python and other high-level languages; this paper will hereafter use the term ``class'' to refer to both low-level structs and high-level classes.)

The LALSuite libraries provide extensive, well-tested routines for gravitational-wave data analysis, in particular for searches for binary black holes and binary neutron stars, which have been carefully vetted by members of the LIGO Scientific Collaboration and Virgo Collaboration.
These include the template signal waveforms for such events, as predicted by general relativity, which tend to be complicated mathematical expressions~\cite[e.g.][]{BuonEtAl2003:DtTFGrvWFSBBlInNnsC} which are time-consuming to implement and verify.
More recent gravitational-wave data analysis software has sought to take advantage of the ease of development and extensive package library of Python; without access to LALSuite routines, however, developers would have faced a significant additional burden in re-implementing and re-verifying the routines in Python.

This paper describes SWIGLAL, which provides Python and Octave interfaces to the libraries provided by LALSuite.
These interfaces have enabled modern gravitational-wave data analysis software to benefit from the advantages of programming in high-level languages, while retaining access to the trusted code base and computational efficiency of the LALSuite code base.

\begin{table}
  \begin{tabular*}{\textwidth}{l @{\extracolsep{\fill}} r @{\extracolsep{\fill}} r @{\extracolsep{\fill}} r @{\extracolsep{\fill}} r @{\extracolsep{\fill}} r}
    \hline
                  & Constants & Variables & Functions & Classes & LOC  \\
    \hline
    LAL(Support)  & 712       & 64        & 1745      & 182     & 38k  \\
    LALBurst      & 9         & 4         & 23        & 2       & 2k   \\
    LALFrame      & 54        & 4         & 254       & 12      & 7k   \\
    LALInference  & 45        & 41        & 416       & 22      & 28k  \\
    LALInspiral   & 204       & 5         & 358       & 58      & 33k  \\
    LALMetaio     & 55        & 4         & 51        & 18      & 3k   \\
    LALPulsar     & 156       & 9         & 623       & 148     & 32k  \\
    LALSimulation & 209       & 42        & 714       & 12      & 92k  \\
    \hline
    Total         & 1444      & 173       & 4184      & 454     & 236k \\
    \hline
\end{tabular*}
\caption{\label{tab:swiglal-exports}
  Number of constants, variables, functions, and classes exported by the SWIGLAL interfaces to the libraries of LALSuite, version 6.67, and an estimate of the total (non-blank, non-comment) lines of C code (LOC) of each library.
  Note that SWIGLAL provides a single interface to the LAL and LALSupport libraries, which are therefore counted together.
}
\end{table}

\section{Software Description and Illustrative Examples}\label{sec:software-description}

Generation of the SWIGLAL interface uses SWIG (Simplified Wrapper and Interface Generator;~\cite{swig}), a software development tool.
SWIG parses the header files of a C/C++ library and identifies the symbols the library exports.
It then generates the wrapper code required to interface the library with a variety of high-level languages, including Python and Octave.
Because it takes C/C++ header files directly as input, SWIG does not require additional code to be written specifically for each exported symbol.
Given the large number of symbols exported by LALSuite (Table~\ref{tab:swiglal-exports}), the automation provided by SWIG relieves LALSuite developers of a significant maintenance burden.
SWIG wrapper code can be further customised by adding \emph{directives} which modify the SWIG-generated wrapper code.
For example, specific directives can be applied to every class in order to add constructors and destructors.
SWIG does not, however, provide a general framework for automating the application of many directives to arbitrary classes of symbols.
To fully automate interface generation, SWIGLAL runs SWIG twice: first as a simple C/C++ header parser, then as an wrapper code generator.
The workflow is as follows:

\begin{enumerate}

\item For each LALSuite library, SWIGLAL generates a basic SWIG interface which simply incorporates all the C header files provided by that library.

\item The basic SWIG interface is input to SWIG with its \texttt{-xml} option, which generates an XML file containing a syntax tree of all symbols exported by the LALSuite library headers.

\item The XML syntax tree is input to a custom Python script, \generateswigiface.
  It parses the XML syntax tree, gathers information about the exported symbols, and generates the full SWIG interface, which augments the basic interface with additional SWIG directives to implement desired functionalities.

\item The full SWIG interface is input to SWIG with its \texttt{-python} or \texttt{-octave} options to generate wrapper code for Python or Octave respectively, which are then compiled into dynamically loadable modules.
  Python modules are loaded using the \lstinline[language=Python]!import! directive; Octave modules are loaded by simply calling the name of the library, e.g.\ ``\lstinline[language=Octave]!lal;!'' for the LAL library.

\end{enumerate}

\begin{figure}
\centering
\begin{lstlisting}
typedef struct tagREAL4Vector {
#ifdef SWIG
    SWIGLAL(ARRAY_STRUCT_1D(REAL4Vector, REAL4, data, UINT4, length));
#endif
    UINT4 length;
    REAL4 *data;
} REAL4Vector;
\end{lstlisting}
\caption{\label{fig:swiglal-macro-example}
  Example usage of the \lstinline!SWIGLAL()! macro in the wrapping of the LAL class \lstinline!REAL4Vector!.
  The \lstinline!ARRAY_STRUCT_1D()! macro exposes the ``\lstinline!data!'' field of the \lstinline!REAL4Vector! class as a native scripting-language array of length ``\lstinline!length!''.
}
\end{figure}

The workflow is implemented as a collection of macros and build rules in the GNU Autoconf/Automake build system used by LALSuite.
Autoconf macros perform configuration tasks, e.g.\ finding a compatible version of the \swig\ binary, and determining the C/C++ preprocessor/compiler/linker flags needed to build the Python/Octave modules.
Automake macros implement the workflow to build the basic and full SWIG interfaces, and the Python/Octave modules, as described above.

A key design objective of SWIGLAL is that the interfaces should resemble and behave, in the supported high-level language, as close to native code written in that language as possible.
To that end, SWIGLAL provides a library of custom SWIG directives which modify the wrapper code to mediate between the expected behaviour of native Python/Octave code and the semantics of the C-language LALSuite libraries.
The interface file \SWIGCommon\ provides common directives used in all languages, while \SWIGPython\ and \SWIGOctave\ provide directives specific to the Python and Octave interfaces respectively.
Each LALSuite library may also provide library-specific directives.

It is also sometimes necessary to add SWIG directives directly to the C header files, in order to further modify the wrapper code for particular functions or classes.
SWIGLAL provides numerous macros, defined in \SWIGCommon\, which are then added to the C header files within \lstinline!#ifdef SWIG!~\dots~\lstinline!#endif! blocks and wrapped in a common macro, \lstinline!SWIGLAL()!; Figure~\ref{fig:swiglal-macro-example} shows an example usage.
This approach keeps SWIG-related code added to the C header files as succinct as possible.
Figure~\ref{fig:swiglal-macro-example} provides an example: the extensive code required to expose the LAL \lstinline!REAL4Vector! class as a native scripting-language array is hidden within the \lstinline!ARRAY_STRUCT_1D()! macro.

The remainder of this section describes some of the issues encountered in fulfilling the objective of the SWIGLAL interface to closely resemble native Python/Octave code, and how those issues are addressed.

\subsection{Class constructors and destructors}

\begin{subfigures}

\begin{figure}
\centering
\begin{lstlisting}
%extend tagLIGOTimeGPS {
  tagLIGOTimeGPS() {
    return %swiglal_new_instance(struct tagLIGOTimeGPS);
  }
  tagLIGOTimeGPS(const struct tagLIGOTimeGPS *src) {
    return %swiglal_new_copy(*src, struct tagLIGOTimeGPS);
  }
  ~tagLIGOTimeGPS() {
    %swiglal_struct_call_dtor(XLALFree, $self);
  }
}
\end{lstlisting}
\caption{\label{fig:swiglal-struct-extend-LTP}
  Example expansion of the \lstinline!\%swiglal_struct_extend()! macro for the LAL class \lstinline!LIGOTimeGPS!.
  This class contains only static fields, and so SWIGLAL provides a constructor, copy constructor, and destructor for this class.
  The SWIG \lstinline!\%extend! directive adds methods to an existing class; methods named after the class are interpreted as constructors, while methods named after the class with the prefix ``\lstinline!\~!'' are interpreted as destructors.
  The \lstinline!\%swiglal_new_instance()! macro allocates a new \lstinline!LIGOTimeGPS! instance using \lstinline!XLALCalloc()!; the \lstinline!\%swiglal_new_copy()! macro creates a copy of an existing \lstinline!LIGOTimeGPS! instance; and the \lstinline!\%swiglal_struct_call_dtor()! macro calls the destructor function \lstinline!XLALFree()!.
}
\end{figure}

\begin{figure}
\centering
\begin{lstlisting}
%extend tagREAL4Vector {
  ~tagREAL4Vector() {
    %swiglal_struct_call_dtor(XLALDestroyREAL4Vector, $self);
  }
}
\end{lstlisting}
\caption{\label{fig:swiglal-struct-extend-R4V}
  Example expansion of the \lstinline!\%swiglal_struct_extend()! macro for the LAL \lstinline!REAL4Vector! class.
  Since this class points in dynamically-allocated memory in its ``\lstinline!data!'' field (Figure~\ref{fig:swiglal-macro-example}), only the destructor is provided, which calls the destructor function \lstinline!XLALDestroyREAL4Vector()!.
}
\end{figure}

\end{subfigures}

LALSuite classes can be separated into two groups, based on their memory requirements.
Classes which contain only static fields, and do not point to dynamically-allocated memory, can be straightforwardly allocated and freed with \lstinline!XLALMalloc()!/\lstinline!XLALCalloc()! and \lstinline!XLALFree()!\footnote{These are LALSuite's equivalents to the C functions \lstinline!malloc()!/\lstinline!calloc()! and \lstinline!free()!, but which also provide optional memory debugging features.}.
For classes which point to dynamically-allocated memory, custom constructor and destructor functions are provided; they are generally named after the class prefixed with ``\lstinline!XLALCreate!\dots'' and ``\lstinline!XLALDestroy!\dots''.
The SWIGLAL \generateswigiface\ script determines to which group each LALSuite class belongs, by using the XML parse tree to determine if a destructor ``\lstinline!XLALDestroy!\dots'' exists for a particular class.
The script then outputs calls to the macro \lstinline!%swiglal_struct_extend()! as part of the full SWIG interface.
Figures~\ref{fig:swiglal-struct-extend-LTP} and~\ref{fig:swiglal-struct-extend-R4V} show two examples of the expansion of \lstinline!%swiglal_struct_extend()! for a class with only static fields (\lstinline!LIGOTimeGPS!) and a class with dynamically-allocated memory (\lstinline!REAL4Vector!).
The provision of correct destructors is necessary to free the user from manual memory management, which high-level languages are expected to handle.
The provision of constructors for classes with static fields provides methods for creating new classes from high-level languages without access to low-level memory functions like \lstinline!XLALMalloc()!.

\subsection{Memory ownership paradigms}

\begin{subfigures}

\begin{figure}
\centering
\begin{lstlisting}
typedef struct tagREAL4TimeSeries {
    CHAR name[LALNameLength];
    LIGOTimeGPS epoch;
    REAL8 deltaT;
    REAL8 f0;
    LALUnit sampleUnits;
    REAL4Vector *data;
} REAL4TimeSeries;
REAL4TimeSeries *XLALCreateREAL4TimeSeries(const CHAR *name, const LIGOTimeGPS *epoch, REAL8 f0, REAL8 deltaT, const LALUnit *sampleUnits, size_t length);
void XLALDestroyREAL4TimeSeries(REAL4TimeSeries *series);
\end{lstlisting}
\caption{\label{fig:memory-ownership-C}
  Illustration of memory ownership tracking in SWIGLAL: Definition of the LAL \lstinline!REAL4TimeSeries! class.
  The ``\lstinline!data!'' field of this class points to an instance of the \lstinline!REAL4Vector! class.
  The \lstinline!XLALCreateREAL4TimeSeries()! function allocates memory for a new \lstinline!REAL4TimeSeries! instance, and for a new \lstinline!REAL4Vector! instance which is pointed to by the ``\lstinline!data!'' field.
  The \lstinline!XLALDestroyREAL4TimeSeries()! function destroys both the \lstinline!REAL4TimeSeries! instance and the pointed-to \lstinline!REAL4Vector! instance.
}
\end{figure}

\begin{figure}
\centering
\begin{lstlisting}[language=Python,numbers=left]
>>> import lal
>>> ts = lal.CreateREAL4TimeSeries("timeseries", 1234567890.0, 0, 1./100, lal.VoltUnit, 10)
>>> ts.data.data = range(0,10)
>>> print(ts.data.data)
[0. 1. 2. 3. 4. 5. 6. 7. 8. 9.]
>>> ts_data = ts.data
>>> del ts
>>> print(ts_data.data)
[0. 1. 2. 3. 4. 5. 6. 7. 8. 9.]
>>> del ts_data
\end{lstlisting}
\caption{\label{fig:memory-ownership-Python}
  Illustration of memory ownership tracking in SWIGLAL: Example usage in Python. The user creates a new \lstinline!REAL4TimeSeries! instance at line~2, and assigns values to the data array pointed to by the \lstinline!REAL4Vector! instance in lines~3 and~4.
  The user stores a reference to the ``\lstinline!data!'' member of the \lstinline!REAL4TimeSeries! instance in line~5, and attempts to delete the \lstinline!REAL4TimeSeries! instance in line~6 using the Python \lstinline!del! operator.
  This does not, however, trigger an immediate call to \lstinline!XLALDestroyREAL4TimeSeries()!, since SWIGLAL knows that the user retains a reference to the \lstinline!REAL4Vector! instance in the variable ``\lstinline!ts_data!''.
  The data contained in the \lstinline!REAL4Vector! instance therefore remains accessible (line~8), and \lstinline!XLALDestroyREAL4TimeSeries()! is called only when ``\lstinline!ts_data!'' is destroyed (line~9).
}
\end{figure}

\end{subfigures}

LALSuite assumes that all class instances are referred to exactly once.
When a class instance is destroyed, all dynamically-allocated memory associated with the instance is freed, including any instances of other classes that are pointed to by the parent instance; put another way, the parent instance ``owns'' the memory of the child instances it points to.
High-level languages, however, allow multiple references to be taken to a particular class instance, and memory is only freed once no references to that instance remain.
Class instances are responsible for freeing their own memory, but do not ``own'' the memory of any instances of other classes they point to.

Figures~\ref{fig:memory-ownership-C} and~\ref{fig:memory-ownership-Python} illustrate how the tension between these different paradigms of memory ownership could potentially cause problems.
The LAL \lstinline!REAL4TimeSeries! class contains a pointer to an instance of the \lstinline!REAL4Vector! class\footnote{Strictly speaking, \lstinline!REAL4TimeSeries! is defined with a pointer to \lstinline!REAL4Sequence!, a synonym for \lstinline!REAL4Vector!.}, and its constructor and destructor functions create and destroy all dynamic memory associated with a \lstinline!REAL4TimeSeries! instance, including the \lstinline!REAL4Vector! pointer (Figure~\ref{fig:memory-ownership-C}).
In Python, however, the \lstinline!REAL4TimeSeries! and \lstinline!REAL4Vector! instances have no parent--child relationship; the user is free to create a \lstinline!REAL4TimeSeries! instance (Figure~\ref{fig:memory-ownership-Python}, line~2), store a reference to the \lstinline!REAL4Vector! instance it points to [line~6], then delete the \lstinline!REAL4TimeSeries! instance [line~7] and assume the \lstinline!REAL4Vector! instance will continue to be valid [line~8].
This is incompatible with the LALSuite memory ownership model, which would destroy the \lstinline!REAL4Vector! instance along with the \lstinline!REAL4TimeSeries! that pointed to it, corrupting the reference stored to the \lstinline!REAL4Vector! by the user.

To resolve this tension, the SWIGLAL interface implements a system which tracks the memory ownership relationship between instances.
In Figure~\ref{fig:memory-ownership-Python}, line~6, SWIGLAL modifies the wrapper code for the ``\lstinline!data!'' field of ``\lstinline!ts!'' to record that the \lstinline!REAL4Vector! instance ``\lstinline!ts.data!'', assigned to ``\lstinline!ts_data!'', is owned by the \lstinline!REAL4TimeSeries! instance.
This record is stored in an associative array called the \emph{parent map}.
The parent map also records a reference count of the number of times ``\lstinline!ts.data!'' has been accessed.
Then, in line~7, the Python \lstinline!del! operator is called on ``\lstinline!ts!'', which would normally immediately call the \lstinline!REAL4TimeSeries! destructor; here SWIGLAL intervenes to check whether ``\lstinline!ts!'' exists in the parent map, i.e.\ whether it owns the memory of another class instance.
Since ``\lstinline!ts!'' owns the memory of ``\lstinline!ts.data!'', the destructor function \lstinline!XLALDestroyREAL4TimeSeries()! is not called, and so the memory allocated for the \lstinline!REAL4Vector! instance stored by ``\lstinline!ts_data!'' is not destroyed.
Finally, in line~10, the Python \lstinline!del! operator is called on ``\lstinline!ts_data!''; here SWIGLAL checks who owns the memory of ``\lstinline!ts_data!'' (i.e. the original ``\lstinline!ts!'' object) and whether there are any outstanding references to that memory.
Since both ``\lstinline!ts!'' and ``\lstinline!ts_data!'' have been destroyed, it is safe for SWIGLAL to call the now call the destructor function \lstinline!XLALDestroyREAL4TimeSeries()! for the \lstinline!REAL4TimeSeries! instance created in line~2.

The SWIGLAL memory ownership tracking system, combined with the native reference counting of objects in Python and Octave, completely frees the user from any manual memory management, as is appropriate for a high-level language, while respecting the LALSuite memory management paradigm.
Memory allocated by LALSuite functions is only freed once it is no longer used, and conversely is retained only as long as needed, thus minimising memory usage.

\subsection{Fixed-length and dynamically-sized arrays}

Gravitational-wave data analysis frequently involves operations on large time- and/or frequency-domain data series, and LALSuite provides many functions and classes to represent such data, such the \lstinline!REAL4Vector! (Figure~\ref{fig:swiglal-macro-example}) and \lstinline!REAL4TimeSeries! (Figure~\ref{fig:memory-ownership-C}) classes.
Such data should be accessible from within the SWIGLAL interface as native array objects, and in an efficient manner without copying of data between the C class instance and its high-level language representation.

SWIGLAL provides several typemaps for converting numerical arrays to/from native array objects; for Python, NumPy~\cite{numpy} arrays are used, while for Octave the native matrix type (or subclasses thereof) are used.
For fixed-length C arrays, SWIGLAL supports both one- and two-dimensional arrays; typemaps are provided for both function arguments and C structure fields.
Dynamically-allocated arrays are typically implemented as specific classes in LALSuite; SWIGLAL provides directives which are added to those classes to provide the type conversion.
For the \lstinline!REAL4Vector! class, for example (Figure~\ref{fig:swiglal-macro-example}), the \lstinline!ARRAY_STRUCT_1D()! macro modifies the wrapper code for the ``\lstinline!data!'' field, so that e.g.\ in Python it accepts any valid sequence of floating-point numbers on assignment, and exposes the ``\lstinline!data!'' field as a NumPy array~\cite{numpy-array} view which directly accesses the underlying C memory.

Some LALSuite array classes store only array data, and nothing else: \lstinline!REAL4Vector! (Figure~\ref{fig:swiglal-macro-example}) is such a class, while \lstinline!REAL4TimeSeries! (Figure~\ref{fig:memory-ownership-C}) contains additional fields.
SWIGLAL provides additional typemaps for pure-array classes such as \lstinline!REAL4Vector! so that functions can accept both class instances and native array objects as arguments.
For example, the Python interface to a function which takes a \lstinline!REAL4Vector! instance as an argument will also accept a NumPy array of the appropriate type.

\subsection{Example: extract strain data at time of GW~150914}

\begin{figure}
\centering
\includegraphics[width=\textwidth]{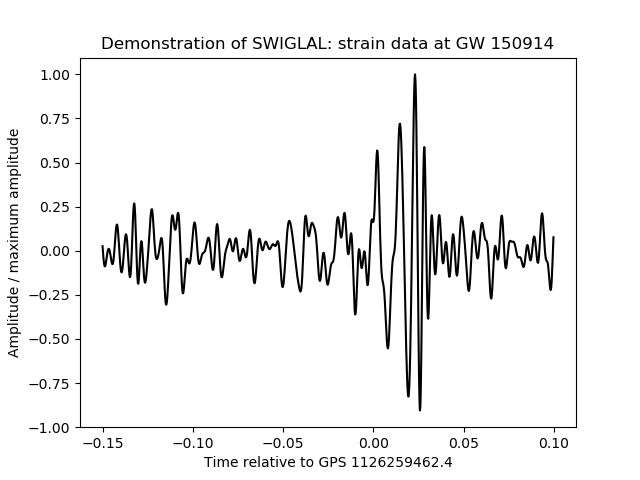}
\caption{\label{fig:example/example}
Whitened, band-pass-filtered strain data from the LIGO Hanford detector at the time of the gravitational-wave event \emph{GW~150914}, as output by the example Python script listed in the Appendix.
}
\end{figure}

Figure~\ref{fig:example/example} shows the output of an example Python script, listed in the Appendix, which extracts the strain data at the time of the first detected gravitational wave event \emph{GW~150914}~\cite[cf. Figure~1 of~][]{LIGOVirg2016:ObsGrvWvBnBHMr}.
The script reads in strain data from the LIGO Hanford detector~\cite{LIGO2015:AdvLIG} at the time of the event, available from~\cite{LIGOVirg2019:ODFrScObsRAdvLAdV}; whitens and band-pass-filters the data so that the event is clearly visible; and plots the processed strain data in the vicinity of the event.
The script is \emph{not} intended as an example of best-practise signal processing for gravitational-wave data analysis, but as an illustration of what may be accomplished in 35 lines of Python code, by harnessing the power of LALSuite routines through the SWIGLAL interface.

\section{Impact}\label{sec:impact}

\begin{table}
  \begin{tabular*}{\textwidth}{l @{\extracolsep{\fill}} r @{\extracolsep{\fill}} r @{\extracolsep{\fill}} r @{\extracolsep{\fill}} r@{\extracolsep{\fill}} r}
    \hline
                  & Sources & Constants & Variables & Functions (LOC) & Classes                        \\
    \hline
    \multicolumn{6}{c}{LALSuite, version 6.67~\cite{lalsuite}. LOC: C $\sim 95\%$, Python $\sim 5$\%.} \\
    LAL(Support)  & 71/222  & 25        & 3         & 32 (3k)         & 18                             \\
    LALBurst      & 39/222  & 0         & 0         & 1 (0.5k)        & 0                              \\
    LALFrame      & 8/222   & 0         & 0         & 20 (4k)         & 4                              \\
    LALInference  & 27/222  & 7         & 3         & 24 (93k)        & 4                              \\
    LALInspiral   & 7/222   & 0         & 0         & 5 (1k)          & 1                              \\
    LALMetaio     & 8/222   & 0         & 0         & 0 (0)           & 2                              \\
    LALPulsar     & 10/222  & 9         & 0         & 43 (16k)        & 21                             \\
    LALSimulation & 20/222  & 0         & 0         & 40 (60k)        & 0                              \\
    \hline
    \multicolumn{6}{c}{PyCBC, version 1.15.4~\cite{pycbc-1.15.4}. LOC: Python $\sim 100$\%.}           \\
    LAL(Support)  & 41/309  & 16        & 0         & 32 (3k)         & 20                             \\
    LALFrame      & 2/309   & 2         & 0         & 25 (6k)         & 1                              \\
    LALPulsar     & 1/309   & 1         & 0         & 5 (2k)          & 2                              \\
    LALSimulation & 14/309  & 4         & 1         & 50 (61k)        & 1                              \\
    \hline
    \multicolumn{6}{c}{GstLAL, version 1.5.1~\tablefootnote{Includes the packages:
    GstLAL Ugly, version 1.6.6, GstLAL Inspiral, version 1.6.9; GstLAL Calibration, version 1.2.11,
    GstLAL Burst, version 0.2.0~\cite{gstlal-1.5.1-etal}.}. LOC: C $\sim 52$\%, Python $\sim 48$\%.}   \\
    LAL(Support)  & 41/117  & 6         & 1         & 16 (2k)         & 11                             \\
    LALSimulation & 9/117   & 0         & 0         & 7 (59k)         & 0                              \\
    \hline
    \multicolumn{6}{c}{Bilby, version 0.6.5~\cite{bilby-0.6.5}. LOC: Python $\sim 100$\%.}             \\
    LAL(Support)  & 8/83    & 3         & 0         & 5 (1k)          & 8                              \\
    LALSimulation & 5/83    & 0         & 0         & 18 (59k)        & 0                              \\
    \hline
    \multicolumn{6}{c}{GWpy, version 1.0.1~\cite{gwpy-1.0.1}. LOC: Python $\sim 100$\%.}               \\
    LAL(Support)  & 10/267  & 10        & 8         & 3 (1k)          & 4                              \\
    LALFrame      & 2/267   & 0         & 0         & 9 (3k)          & 2                              \\
    \hline
    \multicolumn{6}{c}{PyFstat, version 1.3~\cite{PyFstat-1.3}. LOC: Python $\sim 100$\%.}             \\
    LAL(Support)  & 6/28    & 8         & 1         & 5 (1k)          & 6                              \\
    LALPulsar     & 6/28    & 10        & 1         & 22 (14k)        & 21                             \\
    \hline
    \multicolumn{6}{c}{CWInPy, version 0.2.1~\cite{cwinpy-0.2.1}. LOC: Python $\sim 100$\%.}           \\
    LAL(Support)  & 3/42    & 2         & 0         & 4 (1k)          & 4                              \\
    LALInference  & 14/42   & 0         & 1         & 0 (0)           & 0                              \\
    LALPulsar     & 2/42    & 0         & 0         & 1 (1k)          & 0                              \\
    LALSimulation & 1/42    & 0         & 0         & 9 (1k)          & 0                              \\
    \hline
    \multicolumn{6}{c}{OctApps, version 0.2~\cite{octapps-0.2}. LOC: Octave $\sim 100$\%.}             \\
    LAL(Support)  & 9/243   & 7         & 0         & 9 (1k)          & 5                              \\
    LALPulsar     & 7/243   & 29        & 1         & 22 (13k)        & 20                             \\
    \hline
\end{tabular*}
\caption{\label{tab:swiglal-usage}
  Usage of the SWIGLAL interfaces by LALSuite itself, and by the PyCBC, GstLAL, Bilby, GWpy, PyFstat, CWInPy, and OctApps packages.
  The header for each table section gives the package name and version, and the percentage of (non-blank, non-comment) lines of code (LOC) written in C, Python, and/or Octave.
  The columns give the number of source files (out of the total in each package) which reference the SWIGLAL interfaces of each LALSuite library, as well as the number of distinct constants, variables, functions, and classes exported by the SWIGLAL interfaces that are referenced by each package.
  For functions, an estimate of the total (non-blank, non-comment) lines of C code (LOC) represented, including nested calls, is given in parentheses.
  Note that SWIGLAL provides a single interface to the LAL and LALSupport libraries, which are therefore counted together.
}
\end{table}

Table~\ref{tab:swiglal-usage} show the usage of the SWIGLAL interfaces by Python code within LALSuite itself, and by seven other gravitational-wave data analysis packages.
The table gives, for each LALSuite library: the number of source files (out of the package total) which reference the SWIGLAL interface for that library (e.g.\ by importing the interface in Python using ``\lstinline[language=Python]!import!''), and the number of distinct symbols referred to by the package.
The table also lists an estimate of the total lines of C code represented by the LALSuite functions referenced from each package; the estimates include any nested calls to other LALSuite functions.
Python code within LALSuite is a substantial user of SWIGLAL, in terms of source files ($\sim 3$--30\%), and lines of C code utilised ($\sim 180$k).

The PyCBC~\cite{AlleEtAl2012:FIAlDtGrvWInsCB,Alle2005:2TmfDscGrWvDt,NitzEtAl2017:DtBCmpMGrvWUnISnPS,DalCEtAl2014:ImpSAlNStHSAGBGrvWDt} and GstLAL~\cite{gstlal-1,gstlal-2} data analysis packages were used in the first detections of gravitational waves~\cite{LIGOVirg2016:ObsGrvWvBnBHMr,LIGOVirg2017:GWObsGrvWBNtSIn}.
PyCBC makes use of the LAL library in over 10\% of its source files, mostly for manipulating time- and frequency-domain data series.
It uses 25 functions from LALFrame to read and write gravitational wave data in the standard Frame format~\cite{LIGOVirg2009:SpcCmDFFIntGrWDt} produced by gravitational-wave observatories.
It uses 50 functions from LALSimulation to generate template waveforms for matched filtering of the gravitational-wave data.
It uses a few functions from LALPulsar for template bank generation~\cite{Wett2014:LTmPlcChASrGrvP}.
The total lines of LALSuite code utilised by PyCBC is $\sim 72$k.
While primarily written in C, GstLAL uses 16 functions from the LAL library to manipulate time- and frequency-domain data, and compute the geocentric time delay to the gravitational-wave observatories, from Python scripts.
It uses 7 functions from LALSimulation to generate template waveforms in Python.
The total lines of LALSuite code utilised by GstLAL from Python is $\sim 61$k.

Bilby~\cite{AshtEtAl2019:BUsrByInLGrvAs} aims to be a user-friendly package for Bayesian inference for use in gravitational-wave data analysis~\cite[e.g.][]{2020arXiv200101761T}.
It accesses LALSuite through the SWIGLAL interfaces in $\sim 5$--10\% of its source files.
The LAL library is used to handle time- and-frequency domain gravitational-wave data, and LALSimulation is used to generate template waveforms for computing the Bayesian likelihood function.
A total of $\sim 60$k lines of LALSuite code are utilised.

GWpy~\cite{gwpy} is a general package for easily accessing, visualising, and studying gravitational-wave data.
It makes use of the LAL and LALFrame libraries, primarily for manipulating gravitational-wave data in the Frame format, in about $\sim 5$\% of its source files.
It uses $\sim 4$k lines of LALSuite code in total.

PyFstat~\cite{AshtEtAl2018:SmcGlCntSrMt}, CWInPy~\cite{cwinpy-manual}, and OctApps~\cite{WettEtAl2018:OcLOFnCntGrvDAn} are data analysis packages focused on the search for \emph{continuous} gravitational waves from rapidly rotating neutron stars; this class of gravitational wave signals has not yet been detected.
PyFstat uses LAL and LALPulsar (in $\sim 20$\% of its source files) to compute the $\mathcal{F}$-statistic~\cite{JaraEtAl1998:DAnGrvSgSpNSSDtc}, a standard data analysis routine for continuous gravitational wave searches.
CWInPy uses a few LALSuite routines to e.g.\ handle frequency-domain data, convert between time standards, and access properties of the gravitational-wave observatories.
OctApps uses routines, predominately from LALPulsar, to compute the $\mathcal{F}$-statistic and its associated parameter space metric~\cite{Wett2015:PrmMASmSrGrvPl} for designing continuous gravitational wave searches.
Both PyFstat and OctApps use $\sim 14$--$15$k lines of LALSuite code, which CWInPy uses $\sim 3$k.

\section{Conclusions}\label{sec:conclusions}

LALSuite is an important, well-tested component of the gravitational-wave data analysis software stack.
SWIGLAL makes innovative use of SWIG to provide automatically-generated interfaces to LALSuite for Python and Octave, with an emphasis on modelling native code behaviour in those languages.
The interfaces have facilitated the development of modern gravitational-wave data analysis software written in Python, in particular PyCBC which was used in the first discovery of gravitational waves.
The extensive use of the interfaces by a wide variety of Python and Octave packages for gravitational-wave data analysis demonstrates the impact and usefulness of SWIGLAL.

\section{Conflict of Interest}

We wish to confirm that there are no known conflicts of interest associated with this publication and there has been no significant financial support for this work that could have influenced its outcome.

\section*{Acknowledgements}

I am grateful to Adam Mercer for help integrating and maintaining SWIGLAL within LALSuite, and to Kipp Cannon, Jolien Creighton, Nickolas Fotopoulos, Duncan Macleod, and Leo Singer for contributing improvements and bug fixes.
I thank David Keitel and Duncan Macleod for helpful comments on the manuscript.
This work was supported by the Australian Research Council through project number CE170100004, and by the Max Planck Society.
This paper has document number LIGO-P2000094.

\appendix

\section*{Appendix: Example Python script to extract strain data at time of GW~150914}

\begin{lstlisting}[language=Python]
import lal
import lalframe as lalfr
import numpy as np
import matplotlib.pyplot as plt

# read strain data at time of GW 150914
# - frame file downloaded from https://www.gw-openscience.org/
frame_file = lalfr.FrFileOpenURL("./H-H1_GWOSC_4KHZ_R1-1126259447-32.gwf")
gw_strain = lalfr.FrFileReadREAL8TimeSeries(frame_file, "H1:GWOSC-4KHZ_R1_STRAIN", 0)

# compute average power spectral density of strain data
psd_segment_len = int(4.0 / gw_strain.deltaT)
psd_window = lal.CreateTukeyREAL8Window(psd_segment_len, 0.5)
fft_plan = lal.CreateForwardREAL8FFTPlan(psd_segment_len, 0)
gw_psd_length = psd_segment_len // 2 + 1
gw_psd = lal.CreateREAL8FrequencySeries("psd", gw_strain.epoch, 0, 0, lal.DimensionlessUnit, gw_psd_length)
lal.REAL8AverageSpectrumWelch(gw_psd, gw_strain, psd_segment_len, psd_segment_len, psd_window, fft_plan)
gw_psd_f = gw_psd.f0 + np.arange(gw_psd.data.length) * gw_psd.deltaF;

# transform strain data to Fourier domain
gw_fourier_length = gw_strain.data.length // 2 + 1
gw_fourier_deltaF = 0.5 / gw_strain.deltaT / (gw_fourier_length - 1)
fft_plan = lal.CreateForwardREAL8FFTPlan(gw_strain.data.length, 0)
gw_fourier = lal.CreateCOMPLEX16FrequencySeries("fourier", gw_strain.epoch, 0, gw_fourier_deltaF, lal.DimensionlessUnit, gw_fourier_length)
gw_fourier_f = gw_fourier.f0 + np.arange(gw_fourier.data.length) * gw_fourier.deltaF;
lal.REAL8ForwardFFT(gw_fourier.data, gw_strain.data, fft_plan)

# whiten strain data in Fourier domain
gw_psd_at_fourier_f = np.interp(gw_fourier_f, gw_psd_f, gw_psd.data.data)
gw_fourier.data.data = gw_fourier.data.data / np.sqrt(gw_psd_at_fourier_f)

# transform whitened strain data back to time domain
fft_plan = lal.CreateReverseREAL8FFTPlan(gw_strain.data.length, 0)
lal.REAL8ReverseFFT(gw_strain.data, gw_fourier.data, fft_plan)

# band-pass filter whitened time series between 50 and 300 Hz
lal.HighPassREAL8TimeSeries(gw_strain, 50, 0.1, 6)
lal.LowPassREAL8TimeSeries(gw_strain, 300, 0.1, 6)

# extract strain data [-0.15,0.10] seconds around GW 150914
time_of_GW150914 = lal.LIGOTimeGPS("1126259462.4")
first_sample = int(((time_of_GW150914 - 0.15) - gw_strain.epoch) / gw_strain.deltaT)
num_samples = int(0.25 / gw_strain.deltaT)
gw_strain = lal.CutREAL8TimeSeries(gw_strain, first_sample, num_samples)
gw_strain_t = float(gw_strain.epoch - time_of_GW150914) + np.arange(gw_strain.data.length) * gw_strain.deltaT

# plot strain data
plt.plot(gw_strain_t, gw_strain.data.data / max(gw_strain.data.data), "k-")
plt.title("Demonstration of SWIGLAL: strain data at GW 150914")
plt.xlabel(f"Time relative to GPS {time_of_GW150914}")
plt.ylabel("Amplitude / maximum amplitude")
plt.show()
\end{lstlisting}


\begin{thebibliography}{10}
\expandafter\ifx\csname url\endcsname\relax
  \def\url#1{\texttt{#1}}\fi
\expandafter\ifx\csname urlprefix\endcsname\relax\def\urlprefix{URL }\fi
\expandafter\ifx\csname href\endcsname\relax
  \def\href#1#2{#2} \def\path#1{#1}\fi

\bibitem{python}
{Python Software Foundation}, \href{https://docs.python.org/reference/}{{Python
  Language Reference}} (2020).
\newline\urlprefix\url{https://docs.python.org/reference/}

\bibitem{octave}
J.~W. Eaton, D.~Bateman, S.~Hauberg, R.~Wehbring,
  \href{https://www.gnu.org/software/octave/doc/}{{GNU Octave manual: a
  high-level interactive language for numerical computations}} (2020).
\newline\urlprefix\url{https://www.gnu.org/software/octave/doc/}

\bibitem{LIGOVirg2016:ObsGrvWvBnBHMr}
B.~P. Abbott, et~al., {Observation of Gravitational Waves from a Binary Black
  Hole Merger}, Physical Review Letters 116~(6) (2016) 061102.
\newblock \href {http://arxiv.org/abs/1602.03837} {\path{arXiv:1602.03837}},
  \href {http://dx.doi.org/10.1103/PhysRevLett.116.061102}
  {\path{doi:10.1103/PhysRevLett.116.061102}}.

\bibitem{LIGOVirg2017:GWObsGrvWBNtSIn}
B.~P. {Abbott}, et~al., {GW170817: Observation of Gravitational Waves from a
  Binary Neutron Star Inspiral}, Physical Review Letters 119~(16) (2017)
  161101.
\newblock \href {http://arxiv.org/abs/1710.05832} {\path{arXiv:1710.05832}},
  \href {http://dx.doi.org/10.1103/PhysRevLett.119.161101}
  {\path{doi:10.1103/PhysRevLett.119.161101}}.

\bibitem{lalsuite}
{LIGO Scientific Collaboration}, {LIGO} {A}lgorithm {L}ibrary - {LALS}uite,
  Free software (GPL) (2018).
\newblock \href {http://dx.doi.org/10.7935/GT1W-FZ16}
  {\path{doi:10.7935/GT1W-FZ16}}.

\bibitem{lalsuite-6.67}
A.~Mercer,
  \href{https://git.ligo.org/lscsoft/lalsuite/-/tags/lalsuite-v6.67}{{Top-level:
  mark as version 6.67}}, GitLab (Jan. 2020).
\newline\urlprefix\url{https://git.ligo.org/lscsoft/lalsuite/-/tags/lalsuite-v6.67}

\bibitem{BuonEtAl2003:DtTFGrvWFSBBlInNnsC}
A.~{Buonanno}, Y.~{Chen}, M.~{Vallisneri}, {Detection template families for
  gravitational waves from the final stages of binary black-hole inspirals:
  Nonspinning case}, Physical Review D 67~(2) (2003) 024016.
\newblock \href {http://arxiv.org/abs/gr-qc/0205122}
  {\path{arXiv:gr-qc/0205122}}, \href
  {http://dx.doi.org/10.1103/PhysRevD.67.024016}
  {\path{doi:10.1103/PhysRevD.67.024016}}.

\bibitem{swig}
D.~M. Beazley, {SWIG: An Easy to Use Tool for Integrating Scripting Languages
  with C and C++}, in: Proceedings of the 4th Conference on USENIX Tcl/Tk
  Workshop, Vol.~4 of TCLTK’96, USENIX Association, USA, 1996, p.~15.

\bibitem{numpy}
T.~E. Oliphant, \href{http://www.numpy.org/}{{Guide to NumPy}} (2006).
\newline\urlprefix\url{http://www.numpy.org/}

\bibitem{numpy-array}
S.~{van der Walt}, S.~C. {Colbert}, G.~{Varoquaux}, {The NumPy Array: A
  Structure for Efficient Numerical Computation}, Computing in Science
  Engineering 13~(2) (2011) 22.
\newblock \href {http://dx.doi.org/10.1109/MCSE.2011.37}
  {\path{doi:10.1109/MCSE.2011.37}}.

\bibitem{LIGO2015:AdvLIG}
J.~Aasi, et~al., {Advanced LIGO}, Classical and Quantum Gravity 32~(7) (2015)
  074001.
\newblock \href {http://arxiv.org/abs/1411.4547} {\path{arXiv:1411.4547}},
  \href {http://dx.doi.org/10.1088/0264-9381/32/7/074001}
  {\path{doi:10.1088/0264-9381/32/7/074001}}.

\bibitem{LIGOVirg2019:ODFrScObsRAdvLAdV}
R.~{Abbott}, et~al., {Open data from the first and second observing runs of
  Advanced LIGO and Advanced Virgo}, arXiv e-prints (2019) arXiv:1912.11716.
\newblock \href {http://arxiv.org/abs/1912.11716} {\path{arXiv:1912.11716}}.

\bibitem{pycbc-1.15.4}
A.~Nitz, et~al., \href{https://doi.org/10.5281/zenodo.3630601}{{gwastro/pycbc:
  PyCBC Release v1.15.4}}, Zenodo (Jan. 2020).
\newline\urlprefix\url{https://doi.org/10.5281/zenodo.3630601}

\bibitem{bilby-0.6.5}
G.~Ashton, \href{https://github.com/lscsoft/bilby/releases/tag/0.6.5}{{0.6.5 --
  Version 0.6.5 release}}, GitHub (Feb. 2020).
\newline\urlprefix\url{https://github.com/lscsoft/bilby/releases/tag/0.6.5}

\bibitem{gwpy-1.0.1}
D.~Macleod, et~al., \href{https://doi.org/10.5281/zenodo.3598469}{gwpy/gwpy:
  1.0.1} (Jan. 2020).
\newblock \href {http://dx.doi.org/10.5281/zenodo.3598469}
  {\path{doi:10.5281/zenodo.3598469}}.
\newline\urlprefix\url{https://doi.org/10.5281/zenodo.3598469}

\bibitem{PyFstat-1.3}
G.~Ashton, D.~Keitel, R.~Prix,
  \href{https://doi.org/10.5281/zenodo.3620861}{Pyfstat-v1.3} (Jan. 2020).
\newline\urlprefix\url{https://doi.org/10.5281/zenodo.3620861}

\bibitem{cwinpy-0.2.1}
M.~Pitkin, \href{https://github.com/cwinpy/cwinpy/releases/tag/v0.2.1}{{v0.2.1
  -- Release for PyPI}}, GitHub (Jan. 2020).
\newline\urlprefix\url{https://github.com/cwinpy/cwinpy/releases/tag/v0.2.1}

\bibitem{octapps-0.2}
K.~Wette, \href{https://github.com/octapps/octapps/releases/tag/v0.2}{{v0.2 --
  Release for Journal of Open Source Software paper acceptance}}, GitHub (Jun.
  2018).
\newline\urlprefix\url{https://github.com/octapps/octapps/releases/tag/v0.2}

\bibitem{gstlal-1.5.1-etal}
P.~Godwin,
  \href{https://git.ligo.org/lscsoft/gstlal/-/blob/b3b89bc/README.md}{{README.md:
  update versions}}, GitLab (Dec. 2019).
\newline\urlprefix\url{https://git.ligo.org/lscsoft/gstlal/-/blob/b3b89bc/README.md}

\bibitem{AlleEtAl2012:FIAlDtGrvWInsCB}
B.~{Allen}, W.~G. {Anderson}, P.~R. {Brady}, D.~A. {Brown}, J.~D.~E.
  {Creighton}, {FINDCHIRP: An algorithm for detection of gravitational waves
  from inspiraling compact binaries}, Physical Review D 85~(12) (2012) 122006.
\newblock \href {http://arxiv.org/abs/gr-qc/0509116}
  {\path{arXiv:gr-qc/0509116}}, \href
  {http://dx.doi.org/10.1103/PhysRevD.85.122006}
  {\path{doi:10.1103/PhysRevD.85.122006}}.

\bibitem{Alle2005:2TmfDscGrWvDt}
B.~{Allen}, {$\chi^2$ time-frequency discriminator for gravitational wave
  detection}, Physical Review D 71~(6) (2005) 062001.
\newblock \href {http://arxiv.org/abs/gr-qc/0405045}
  {\path{arXiv:gr-qc/0405045}}, \href
  {http://dx.doi.org/10.1103/PhysRevD.71.062001}
  {\path{doi:10.1103/PhysRevD.71.062001}}.

\bibitem{NitzEtAl2017:DtBCmpMGrvWUnISnPS}
A.~H. {Nitz}, T.~{Dent}, T.~{Dal Canton}, S.~{Fairhurst}, D.~A. {Brown},
  {Detecting Binary Compact-object Mergers with Gravitational Waves:
  Understanding and Improving the Sensitivity of the PyCBC Search},
  Astrophysical Journal 849~(2) (2017) 118.
\newblock \href {http://arxiv.org/abs/1705.01513} {\path{arXiv:1705.01513}},
  \href {http://dx.doi.org/10.3847/1538-4357/aa8f50}
  {\path{doi:10.3847/1538-4357/aa8f50}}.

\bibitem{DalCEtAl2014:ImpSAlNStHSAGBGrvWDt}
T.~Dal~Canton, A.~H. Nitz, A.~P. Lundgren, A.~B. Nielsen, D.~A. Brown, T.~Dent,
  I.~W. Harry, B.~Krishnan, A.~J. Miller, K.~Wette, K.~Wiesner, J.~L. Willis,
  {Implementing a search for aligned-spin neutron star-black hole systems with
  advanced ground based gravitational wave detectors}, Physical Review D 90~(8)
  (2014) 082004.
\newblock \href {http://arxiv.org/abs/1405.6731} {\path{arXiv:1405.6731}},
  \href {http://dx.doi.org/10.1103/PhysRevD.90.082004}
  {\path{doi:10.1103/PhysRevD.90.082004}}.

\bibitem{gstlal-1}
S.~{Sachdev}, S.~{Caudill}, H.~{Fong}, R.~K.~L. {Lo}, C.~{Messick},
  D.~{Mukherjee}, R.~{Magee}, L.~{Tsukada}, K.~{Blackburn}, P.~{Brady},
  P.~{Brockill}, K.~{Cannon}, S.~J. {Chamberlin}, D.~{Chatterjee}, J.~D.~E.
  {Creighton}, P.~{Godwin}, A.~{Gupta}, C.~{Hanna}, S.~{Kapadia}, R.~N. {Lang},
  T.~G.~F. {Li}, D.~{Meacher}, A.~{Pace}, S.~{Privitera}, L.~{Sadeghian},
  L.~{Wade}, M.~{Wade}, A.~{Weinstein}, S.~{Liting Xiao}, {The GstLAL Search
  Analysis Methods for Compact Binary Mergers in Advanced LIGO's Second and
  Advanced Virgo's First Observing Runs}, arXiv (2019) 1901.08580.
\newblock \href {http://arxiv.org/abs/1901.08580} {\path{arXiv:1901.08580}}.

\bibitem{gstlal-2}
C.~{Messick}, K.~{Blackburn}, P.~{Brady}, P.~{Brockill}, K.~{Cannon},
  R.~{Cariou}, S.~{Caudill}, S.~J. {Chamberlin}, J.~D.~E. {Creighton},
  R.~{Everett}, C.~{Hanna}, D.~{Keppel}, R.~N. {Lang}, T.~G.~F. {Li},
  D.~{Meacher}, A.~{Nielsen}, C.~{Pankow}, S.~{Privitera}, H.~{Qi},
  S.~{Sachdev}, L.~{Sadeghian}, L.~{Singer}, E.~G. {Thomas}, L.~{Wade},
  M.~{Wade}, A.~{Weinstein}, K.~{Wiesner}, {Analysis framework for the prompt
  discovery of compact binary mergers in gravitational-wave data}, Physical
  Review D 95~(4) (2017) 042001.
\newblock \href {http://arxiv.org/abs/1604.04324} {\path{arXiv:1604.04324}},
  \href {http://dx.doi.org/10.1103/PhysRevD.95.042001}
  {\path{doi:10.1103/PhysRevD.95.042001}}.

\bibitem{LIGOVirg2009:SpcCmDFFIntGrWDt}
{LIGO}, {Virgo}, {Specification of a Common Data Frame Format for
  Interferometric Gravitational Wave Detectors}, Tech. Rep. LIGO-T970130-v1,
  VIR-067A-08, LIGO, Virgo (2009).

\bibitem{Wett2014:LTmPlcChASrGrvP}
K.~Wette, {Lattice template placement for coherent all-sky searches for
  gravitational-wave pulsars}, Physical Review D 90 (2014) 122010.
\newblock \href {http://arxiv.org/abs/1410.6882} {\path{arXiv:1410.6882}},
  \href {http://dx.doi.org/10.1103/PhysRevD.90.122010}
  {\path{doi:10.1103/PhysRevD.90.122010}}.

\bibitem{AshtEtAl2019:BUsrByInLGrvAs}
G.~{Ashton}, M.~{H{\"u}bner}, P.~D. {Lasky}, C.~{Talbot}, K.~{Ackley},
  S.~{Biscoveanu}, Q.~{Chu}, A.~{Divakarla}, P.~J. {Easter}, B.~{Goncharov},
  F.~{Hernandez Vivanco}, J.~{Harms}, M.~E. {Lower}, G.~D. {Meadors},
  D.~{Melchor}, E.~{Payne}, M.~D. {Pitkin}, J.~{Powell}, N.~{Sarin}, R.~J.~E.
  {Smith}, E.~{Thrane}, {BILBY: A User-friendly Bayesian Inference Library for
  Gravitational-wave Astronomy}, Astrophysical Journal Supplement 241~(2)
  (2019) 27.
\newblock \href {http://arxiv.org/abs/1811.02042} {\path{arXiv:1811.02042}},
  \href {http://dx.doi.org/10.3847/1538-4365/ab06fc}
  {\path{doi:10.3847/1538-4365/ab06fc}}.

\bibitem{2020arXiv200101761T}
B.~P. {Abbott}, et~al., {GW190425: Observation of a Compact Binary Coalescence
  with Total Mass $\sim 3.4 M_{\odot}$}, arXiv (2020) 2001.01761.
\newblock \href {http://arxiv.org/abs/2001.01761} {\path{arXiv:2001.01761}}.

\bibitem{gwpy}
D.~Macleod, et~al., {GWpy: a python package for gravitational-wave
  astrophysics}, Software X (2020) submitted.

\bibitem{AshtEtAl2018:SmcGlCntSrMt}
G.~{Ashton}, R.~{Prix}, D.~I. {Jones}, {A semicoherent glitch-robust
  continuous-gravitational-wave search method}, Physical Review D 98~(6) (2018)
  063011.
\newblock \href {http://arxiv.org/abs/1805.03314} {\path{arXiv:1805.03314}},
  \href {http://dx.doi.org/10.1103/PhysRevD.98.063011}
  {\path{doi:10.1103/PhysRevD.98.063011}}.

\bibitem{cwinpy-manual}
M.~Pitkin, \href{https://cwinpy.readthedocs.io/en/latest/}{{Documentation for
  CWInPy}} (2019).
\newline\urlprefix\url{https://cwinpy.readthedocs.io/en/latest/}

\bibitem{WettEtAl2018:OcLOFnCntGrvDAn}
K.~Wette, R.~Prix, D.~Keitel, M.~Pitkin, C.~Dreissigacker, J.~T. Whelan,
  P.~Leaci, {OctApps: a library of Octave functions for continuous
  gravitational-wave data analysis}, Journal of Open Source Software 3~(26)
  (2018) 707.
\newblock \href {http://dx.doi.org/10.21105/joss.00707}
  {\path{doi:10.21105/joss.00707}}.

\bibitem{JaraEtAl1998:DAnGrvSgSpNSSDtc}
P.~Jaranowski, A.~Kr\'olak, B.~F. Schutz, {Data analysis of gravitational-wave
  signals from spinning neutron stars: The signal and its detection}, Physical
  Review D 58~(6) (1998) 063001.
\newblock \href {http://arxiv.org/abs/gr-qc/9804014}
  {\path{arXiv:gr-qc/9804014}}, \href
  {http://dx.doi.org/10.1103/PhysRevD.58.063001}
  {\path{doi:10.1103/PhysRevD.58.063001}}.

\bibitem{Wett2015:PrmMASmSrGrvPl}
K.~Wette, {Parameter-space metric for all-sky semicoherent searches for
  gravitational-wave pulsars}, Physical Review D 92~(8) (2015) 082003.
\newblock \href {http://arxiv.org/abs/1508.02372} {\path{arXiv:1508.02372}},
  \href {http://dx.doi.org/10.1103/PhysRevD.92.082003}
  {\path{doi:10.1103/PhysRevD.92.082003}}.

\end{thebibliography}
\end{document}